\documentclass{PoS}

\usepackage{amsmath}
\usepackage{amsfonts}
\usepackage{amssymb}
\usepackage{bbm}
\usepackage{physics}
\usepackage{verbatim}
\usepackage{dsfont}
\usepackage{booktabs}
\usepackage{slashed}
\usepackage{appendix}
\usepackage{epsfig}
\usepackage{color}
\usepackage[table]{xcolor}
\usepackage{graphicx}
\usepackage[]{caption}
\usepackage[listofformat=empty,subrefformat=empty]{subfig}
\usepackage{float}
\usepackage{overpic}
\usepackage{enumerate}
\usepackage{hhline}
\usepackage{multirow}
\usepackage{cite}
\usepackage{xspace}
\usepackage{setspace}

\def\be{\begin{equation}}
\def\ee{\end{equation}}
\def\bea{\begin{eqnarray}}
\def\eea{\end{eqnarray}}
\def\ba{\begin{aligned}}
\def\ea{\end{aligned}}
\def\cL{{\cal L}}

\newcommand{\lp}{\left(}
\newcommand{\rp}{\right)}
\newcommand{\lb}{\left[}
\newcommand{\rb}{\right]}

\title{Weak gravity (and other conjectures) with broken supersymmetry}

\ShortTitle{Weak gravity (and other conjectures) with broken supersymmetry}

 \author{ \speaker{Quentin Bonnefoy}$^{a}$, Emilian Dudas$^b$,
     and Severin L\"ust$^{bc}$\\
     \llap{$^a$}DESY, Notkestra\ss e 85, 22607 Hamburg, Germany\\
     \llap{$^b$}Centre de Physique Th{\'e}orique, CNRS, {\'E}cole Polytechnique, IP Paris, 91128 Palaiseau, France\\
     \llap{$^c$}Institut de Physique Th\'eorique, Universit\'e Paris Saclay, CEA, CNRS, F-91191 Gif sur Yvette, France\\
     E-mail:  \email{quentin.bonnefoy@desy.de}, \email{emilian.dudas@polytechnique.edu},
     \email{severin.luest@polytechnique.edu}}

\abstract{We study the weak gravity conjecture in non-supersymmetric string theory setups. Precisely, those are type I string theory with supersymmetry broken \`a la Scherk-Schwarz and open strings on D branes wrapped around magnetized tori in type II string theory. We compute long-range interactions between identical branes at one-loop and compare them to the weak gravity conjecture for higher-degree forms. In our examples, SUSY breaking generates interactions between branes, which are not anymore BPS, in such a way that the weak gravity conjecture is verified. In  type I with the Scherk-Schwarz mechanism, the tension of the branes is reduced by one-loop quantum effects, so that there are long-range repulsive forces. The correlation of the non-vanishing brane potential with the presence of a running modulus and of possible D branes bound states nicely connects to other swampland conjectures. For magnetized branes in type II strings, we check that non-BPS branes experience a long-range repulsion whenever the open string spectrum is tachyon-free. Ultimately, the role of stringy objects in the discussion makes it compelling to further understand swampland conjectures in strings with broken SUSY, let alone their phenomenological relevance.}

\FullConference{Corfu Summer Institute 2019 "School and Workshops on Elementary Particle Physics and Gravity" (CORFU2019)\\
		31 August - 25 September 2019\\
		Corfu, Greece}

\begin{document}



\section{Introduction}

Swampland conjectures \cite{Vafa:2005ui} constrain the properties of effective field theories (EFTs) that could possibly derive from a full theory of quantum gravity (for a recent review and a complete set of references, see \cite{Palti:2019pca}). Most notably, they concern the validity domain of EFTs \cite{Ooguri:2006in}, their possible global and local symmetries \cite{Banks:2010zn}, the spectrum of gauge theories \cite{ArkaniHamed:2006dz}, the stability of non-supersymmetric AdS spacetimes \cite{Ooguri:2016pdq} and the shape of scalar potentials \cite{Obied:2018sgi,Ooguri:2018wrx}. They introduce both qualitative insights and quantitative criteria: a canonical example of the former is the presence of towers of states \cite{Ooguri:2006in,Heidenreich:2015nta,Heidenreich:2016aqi,Klaewer:2016kiy,Andriolo:2018lvp,Grimm:2018ohb}, whereas the concrete behaviour of such towers in specific limits of the theory illustrate the latter aspect. In most cases, both aspects are tested simultaneously, for instance by studying precise string theory setups or using classical results on black holes. 

Here, we will be mostly interested in the weak gravity conjecture (WGC) \cite{ArkaniHamed:2006dz} in specific string theory setups, but our discussion will feature other conjectures.

\subsection{The WGC for $p$-forms}

The WGC is a statement about how gravity and gauge fields can possibly coexist, and it was originally derived in $4$ dimensions by demanding that charged black holes should be 
able to decay \cite{ArkaniHamed:2006dz}. Balancing charge and energy conservation against the extremality condition, one gets that one of the decay products must have loosely speaking its mass smaller than its gauge charge, ensuring that gravity indeed acts as the weakest force on this state. This statement can be extended to $p$-forms gauge fields $A_p$ in arbitrary $d$ spacetime dimensions \cite{Heidenreich:2015nta}. Indeed, with an Einstein frame Lagrangian 
\be
\cL = \frac{M_P^{d-2}}{2}\int d^d x \sqrt{-g}\lp R-\frac{1}{2}(\nabla\phi)^2\rp-\frac{1}{2e^2}\int d^d x \sqrt{-g} \ e^{-\alpha\phi}F_{p+1}^2 \ ,
\ee
where $\phi$ is a dilaton field and $F_{p+1}\equiv dA_p$, there are black branes solutions carrying a (electric or magnetic) charge under $A_p$. By demanding that they could decay, one obtains that there 
should be a ($p-1$)-dimensional object (hence with a $p$-dimensional wordvolume) with tension $T_p$ charged under $A_p$ with quantized charge $Q_p$, so that
\be
\lp\frac{\alpha^2}{2}+\frac{p(d-p-2)}{d-2}\rp T_p^2 \leq e^2 Q_p^2 M_P^{d-2} \ .
\ee
D branes in superstring setups saturate this inequality due to their BPS nature. This is understood from the fact that mutually BPS D branes do not interact, since their gravitational attraction from the exchange of Neveu-Schwarz-Neveu-Schwarz (NS-NS) closed strings precisely cancels the gauge electric repulsion due to the Ramond-Ramond (R-R) sector. Indeed, in the absence of scalar fields, one could reformulate the weak gravity conjecture as the requirement for the existence of at least one particle or brane for each gauge symmetry such that its effective interaction potential is repulsive \cite{Palti:2017elp,Heidenreich:2019zkl}. In what follows, we compute such interaction potentials in explicit string theory models in order to test if they are repulsive and if the branes obey the weak gravity conjecture.

\subsection{SUSY breaking and the swampland}

The specific models we will study have broken supersymmetry (SUSY). This generates several interesting features for non-trivial tests:  effective brane-brane interactions and the generation of scalar potentials (e.g. runaway potentials for moduli fields\footnote{Runaway potentials are abundant in string theory and this was considered as a serious phenomenological problem in the past \cite{Dine:1985he}. Motivated by the persistent presence of runaway potentials in string theory, it was also recently conjectured in \cite{Agrawal:2018own} that quintessence may be the only realistic outcome of a theory of quantum gravity.}). Although not realistic at the level of the models we discuss here, such potentials could be interpreted as dark energy and shed some light on realizations of the swampland with non-trivial cosmological dynamics, favored by \cite{Danielsson:2018ztv,Obied:2018sgi,Ooguri:2018wrx,Agrawal:2018own}. Brane interactions are as we said previously crucial for the WGC. Thus, it seems important to go beyond the majority of string theory tests of the swampland conjectures, which were done in the context of superstring compactifications.

Arguably, the simplest and best understood way of breaking supersymmetry in string theory is via compactification (with a SUSY breaking scale at the compactification scale). This was first proposed at the field-theory (supergravity) level by Scherk and Schwarz  \cite{Scherk:1979zr}. We will use it in what follows, and it will generate a runaway (space) direction in which one field rolls. While in the decompactification limit supersymmetry is restored and the weak gravity conjecture is marginally satisfied, considering the rolling field at a different value generates brane interactions and thus constraints from the point of view of the weak gravity conjecture. 

We will also explore setups with magnetic fields on D branes, which have the generic effect of breaking supersymmetry and departing from the BPS nature of those branes. 
When magnetic fields are not tuned so that SUSY is broken, branes exerce forces on one another which can also be analyzed from the point of view of the weak gravity conjecture. 

\subsection{Outline}

In section \ref{typeISSsection}, we study type I string theory with supersymmetry breaking by compactification. The latter generates a bulk runaway potential for the radius of the compactified direction, see section \ref{bulkPotential}. We compute in section \ref{d1oneloop} the brane-brane interactions at one-loop and discuss their attractive nature. Specifically, we use D1 brane interactions as a function of the separation in spacetime as a test of the WGC. We find that at short distances and at one-loop there are attractive forces which have a finite limit when the distance goes to zero, whereas at long distances those attractive forces are exponentially suppressed. Since massive (closed strings) fields do not mediate long range interactions, our interpretation is that at this order of perturbation theory the branes still have a charge to mass ratio set by the supersymmetric BPS condition. Thus, arguments beyond one-loop are needed in order to clarify the fate of the weak gravity conjecture in this setup. The way to proceed comes from the identification of the quantum corrections to brane tensions (see section \ref{beyondoneloop}): indeed, the limit of zero distance suggests that the corresponding self-energy can be interpreted as a negative quantum correction to the tension, which will generate an imbalance between gauge and gravitational forces at higher loops, leading to an effective repulsion at large distances consistent with the WGC. The one-loop attractive forces, unsuppressed at small distances, will induce the formation of a finite number of stable bound states of D1 branes which we discuss in section \ref{boundstates}. For very small string coupling, the number of such states can become very large, which is reminiscent of the swampland distance conjecture \cite{Ooguri:2006in}, although we cannot consistently test this idea in our setup.

In section \ref{magneticSection}, we discuss the relationship between SUSY breaking via magnetic fields in compact dimensions and the weak gravity conjecture. We start by a lightning-speed review of SUSY breaking via magnetic fields in toroidal compactifications of field and string theory in section \ref{magneticSUSYbreaking}, and we also discuss the correlation between SUSY and the presence of tachyons in the open string spectrum for magnetic fields in one or two torii. This leads us to consider setups with three magnetic fields in section \ref{tachyonsWGC}, where we recall that SUSY can be broken with a tachyon-free spectrum when the three magnetic fields verify certain conditions. Then, we show that D6 branes in type IIA string theory repel (if far apart) precisely when the tachyons are absent. We also discuss the shape of their potential at any distance, taking into account all the string modes, which is relevant for bound states discussions.

In addition to intermediate summaries at the end of sections \ref{typeISSsection} and \ref{magneticSection}, we present some general conclusions in section \ref{summary}.

\section{Type I string theory with Scherk-Schwarz SUSY breaking}\label{typeISSsection}

We review in this section the results of \cite{Bonnefoy:2018tcp}. We are interested in moduli potentials and brane interactions from SUSY breaking, so we consider type I string theory\footnote{Type I is the unoriented version of type IIB string theory, meaning the theory which emerges when type IIB strings are projected on the subsector which preserves the exchange symmetry of left and right string oscillators - this is called the orientifold projection (together with the addition of an open string sector necessary for the consistency of the theory). Thus, the string worldsheet to sum over in perturbative computations contain unoriented 2D surfaces. The 10D low energy effective theory is $N=1$ supergravity with $SO(32)$ gauge group.} and orientifolds 
which contain the necessary ingredients. The vacuum energy and brane-brane interactions are nicely encoded in one-loop vacuum string amplitudes, which correspond to the contribution of the torus and the Klein bottle for the propagation of closed strings, and of the cylinder and the M\"obius strip for open strings. We implement in this section the breaking of supersymmetry \`a la Scherk-Schwarz. In what follows, all string amplitudes should be multiplied by the factor $1/(4 \pi^2 \alpha')^{d/2}$, where $d$ is the number of noncompact spacetime dimensions.

\subsection{The vacuum energy}\label{bulkPotential}

Much of the physics of the vacuum structure of our setup can be understood from its one-loop approximation. With 10D supersymmetric type I strings, one gets a one-loop contribution from the torus which reads
\be
{\cal T} = \int_{\cal F} \frac{d^2 \tau}{\tau_2^6}\abs{\frac{V_8-S_8}{\eta^8}(\tau)}^2 \ ,
\label{SUSYtorus10D}
\ee
where $\tau$ is the complex parameter of the torus, $\cal F$ the fundamental domain of its modular group $\mathrm{SL}(2,\mathbb{Z})$ and $V_8=\frac{\theta_3^4-\theta_4^4}{2\eta^4},S_8=\frac{\theta_2^4+\theta_1^4}{2\eta^4}$ are $SO(8)$ characters built out of Jacobi theta functions (we refer to \cite{Dudas:2000bn,Angelantonj:2002ct} for notations and conventions which we will not always detail in what follows). Compactifying one dimension on a circle of radius $R$, we get instead
\be
{\cal T} = \int_{\cal F} \frac{d^2 \tau}{\tau_2^{11/2}}\abs{\frac{V_8-S_8}{\eta^8}}^2\Lambda_{m,n}(\tau) \ ,
\label{SUSYtorus9D}
\ee
with $\Lambda_{m,n}=\sum_{m,n}q^{\frac{\alpha'}{4}\lp\frac{m}{R}+\frac{nR}{\alpha'}\rp^2}\overline{q}^{\frac{\alpha'}{4}\lp\frac{m}{R}-\frac{nR}{\alpha'}\rp^2}$ (and $q=e^{2i\pi\tau}$) the sum over Kaluza-Klein (KK) and winding modes (respectively indexed by $m$ and $n$) which replace the contribution of the continuous momentum along the compactified dimension. Finally, introducing the Scherk-Schwarz projection along the compactified dimension\footnote{The idea behind the Scherk-Schwarz mechanism is to generate mass shifts via twisted boundary conditions for a theory defined on a compact manifold. It is most easily illustrated for a field theory on a circle of radius $R$, where one can define boundary conditions as follows: field$(x+2\pi R)$ \ = \ symmetry $\times$ field$(x)$. For instance for a charged scalar field with a 
global $U(1)$ symmetry, we can choose $\phi(x+2\pi R)=e^{i\pi Q}\phi(x)$, with $Q$ the charge operator. The result is that the mass of the KK modes is shifted:
\be
\phi(x)=e^\frac{ixQ}{2R}\sum_me^\frac{imx}{R}\phi_m\implies M_{\phi_m}=\abs{\frac{m}{R}+\frac{Q}{2R}} \ .
\ee
If one implements this kind of shift differently for bosons and fermions, SUSY is broken since the mass degeneracy within a multiplet is lifted. At the level of the string, this is achieved by orbifolding the theory under the symmetry $(-1)^F\delta_{X_9\to X_9+\pi R}$, where $X_9$ is a compactified coordinate of periodicity $2\pi R$ and $(-1)^F$  is the spacetime fermion number.}, which at the level of the string 
theory is a freely-acting orbifold, we obtain
\be
\ba
{\cal T} = \int_{\cal F} \frac{d^2 \tau}{\tau_2^{11/2}} \Bigl\{&  (|V_8|^2 + |S_8|^2) \Lambda_{m,2n} - (V_8  {\bar S}_8 +  S_8  {\bar V}_8  ) \Lambda_{m+1/2,2n}    \\
& +   (|O_8|^2 + |C_8|^2) \Lambda_{m,2n+1} - (O_8  {\bar C}_8 +  C_8  {\bar O}_8  ) \Lambda_{m+1/2,2n+1}      \Bigr\}   \frac{1}{|\eta^8|^2} (\tau) \ .
\label{ss1}
\ea
\ee
Lots of things can be read from this expression. First, the indices of the KK/winding sums indicate clearly that fermions (which correspond to characters entering the loop with a minus sign) and bosons (... with a plus) do not have the same masses anymore: SUSY is broken as announced. Second, one can see from this expression that the (torus contribution to the) vacuum energy is not anymore zero unlike the SUSY case of \eqref{SUSYtorus10D}. Finally, one may notice that the second line has no counterpart in \eqref{SUSYtorus9D}, which comes from the fact that it corresponds to the twisted sector which one must include to preserve the modular invariance of the theory. This sector contains a tower of states starting with a scalar (coming from the character $ |O_8|^2$ above) with the lightest mass given by
\be
m_O^2 = - \frac{2}{\alpha'} + \frac{R^2}{\alpha'^2}   \ . \label{ss01}
\ee
For small radii $R < \sqrt{2 \alpha'}$ this scalar becomes tachyonic, whereas it is very heavy in the opposite limit   $R \gg \sqrt{2 \alpha'}$.  This scalar will be a main actor in the brane-brane interactions at long distances that we discuss later on. 

The full vacuum amplitude is obtained by adding the three other worldsheets of Euler number zero: the Klein bottle in the closed string sector, as well as the annulus and the M\"obius strip in the open one. The open sector is associated to the introduction of $16$ D9 branes, required for the consistency of the theory (namely, by the R-R tadpole conditions). They respectively read
\be
\ba
{\cal K} &= \frac{1}{2} \int_0^{\infty} \frac{d \tau_2}{\tau_2^{11/2}} \frac{V_8-S_8}{\eta^8} (2 i \tau_2) \sum_m e^{- \alpha' \pi \tau_2 \frac{m^2}{R^2}} \ , \\
{\cal A} &=    512 \int_0^{\infty} \frac{d \tau_2}{\tau_2^{11/2}}  \lb\frac{V_8}{\eta^8} P_m -\frac{S_8}{\eta^8}P_{m+1/2} \rb\left(\frac{i \tau_2}{2} \right)  \ , \\
{\cal M} &=  -   16 \int_0^{\infty} \frac{d \tau_2}{\tau_2^{11/2}}  \left[\frac{V_8}{\eta^8}P_{m} - \frac{S_8}{\eta^8} P_{m+1/2}  \right]\left(\frac{i \tau_2}{2} + \frac{1}{2}\right)   \ , 
\ea
\ee  
where $P_m=\Lambda_{m,0}$ and we neglected any Wilson lines on the branes (and we will keep on doing so in what follows, we refer to \cite{Bonnefoy:2018tcp} for more details). From this amplitude, we can write explicitly the scalar potential for the radius $R$. The scalar potential in string theory is minus the partition function, therefore
\be
V (R) = - \left( \frac{1}{2} {\cal T} + {\cal K} +  {\cal A} + {\cal M}\right) \ .
\label{Radiuspotential}
\ee
The potential can be easily estimated in the regime where effective field theory is valid $R \gg \sqrt{2 \alpha'}$. In this limit, string oscillators in all amplitudes and winding states in the torus are very heavy and do not contribute.  We can therefore replace the modular functions by their leading contribution, and perform a Poisson resummation of the Kaluza-Klein sums to turn them into winding sums, so that we get
\be
\ba
&{\cal T} \simeq 128  \frac{R}{\sqrt{\alpha'}} \int _0^{\infty} \frac{d \tau_2}{\tau_2^{6}}   \sum_n \left[ 1- (-1)^n \right]  e^{-   \frac{\pi n^2 R^2}{ \alpha' \tau_2 }}  \ , \\
{\cal A} \simeq 4096  \frac{R}{\sqrt{\alpha'}} \int_0^{\infty} \frac{d \tau_2}{\tau_2^{6}}   \sum_n&\left[ 1- (-1)^n \right] e^{-   \frac{\pi n^2 R^2}{ \alpha' \tau_2 }} \ , \quad {\cal M} =  -  128\frac{R}{\sqrt{\alpha'}} \int_0^{\infty} \frac{d \tau_2}{\tau_2^{6}}    \sum_n\left[ 1- (-1)^n \right]  e^{-   \frac{\pi n^2 R^2}{ \alpha' \tau_2 }}  \ .
\ea
\ee
The Klein bottle is still supersymmetric and does not contribute to the scalar potential, so that we do not include it further. As explained at the beginning of this section, all string amplitudes above should be multiplied by the factor $1/(4 \pi^2 \alpha')^{9/2}$. By including this factor and after a straightforward integration, one gets 
\be
{\cal T}  =  \frac{12}{\pi^{14}R^9} \sum_n \frac{1}{(2n+1)^{10}}  \ , \quad {\cal A} \simeq \frac{384}{\pi^{14}R^9}  \sum_n \frac{1}{(2n+1)^{10}} \ , \quad {\cal M}   \simeq  - \frac{12}{\pi^{14}R^9}  \sum_n \frac{1}{(2n+1)^{10}}  \ ,
\label{oneloopamplitudes}
\ee
which generate a runaway potential, a well-known feature of broken SUSY models. Indeed, the 9D effective potential for the radius in the Einstein frame is of the form
  \be     
 {\cal L} = \frac{1}{2\kappa_9^2R^2} (\partial R)^2 - \frac{ce^{\frac{18\phi}{7}}}{R^9}  \  ,  \label{comp2}
 \ee     
where $\phi$ is the dilaton field, $\frac{1}{\kappa_9^2}$ is the nine-dimensional Planck mass and $-\frac{c}{R^9}$ is obtained when summing the three contributions in (\ref{oneloopamplitudes}), according to (\ref{Radiuspotential}). Supersymmetry is then restored in the  limit $R \to \infty$. Although a runaway potential is expected for a quintessence model, we will not enter here into a phenomenological discussion of such potentials and their viability. In particular, the potential \eqref{Radiuspotential} is not viable for several reasons, for instance because of time dependent coupling constants or because the vacuum energy is negative. Making the potential better behaved\footnote{For instance, it is possible to obtain a positive potential with stable brane configurations \cite{Abel:2018zyt,Abel:2020ldo} without altering significantly the discussion on the weak gravity conjecture below.} is beyond the scope of this discussion, however we expect that the qualitative aspects to be commented on soon could be realized in refined phenomenological models, so we proceed and identify the features of the current setup relevant for the weak gravity conjecture.  

The formulae above can be generalized easily after compactification to four dimensions on five more circles of radii $R_I$, $I = 1, \dots, 5$. The vacuum energy, in the large radii limit, becomes
\be
\ba
{\cal T}  =  \frac{192 V_6 }{\pi^{9}} \sum_{\bf n}&  \frac{1 - (-1)^{n}}{[n^{2} R^2 + n_1^2 R_1^2 + \cdots n_5 R_5^2]^5} \ , \\ 
{\cal A}  \simeq \frac{6144 V_6 }{ \pi^{9}}\sum_{\bf n}  \frac{1 - (-1)^{n}}{ [n^{2} R^2 + n_1^2 R_1^2 + \cdots n_5 R_5^2]^5}&\ , \quad {\cal M}   \simeq  - \frac{48 V_6}{4 \pi^{9}} \sum_{\bf n} \frac{1 - (-1)^{n}}{ [n^{2} R^2 + n_1^2 R_1^2 + \cdots n_5 R_5^2]^5}  \  .   
\ea
\ee
 where $V_6 = \prod_I R_I$.

\subsection{D1 branes interactions}\label{d1oneloop}

We now compute branes self-interactions in the setup discussed in section \ref{bulkPotential}. Specifically, we look at D1 branes which couple to gravity and to the R-R $2$-form and derive their one-loop interaction potential. The usual string theory computation of brane-brane interactions \cite{Polchinski:1995mt,Polchinski:1996na,Bachas:1998rg} can be captured at large separations ($\abs{\Delta \vec x} \gg \sqrt{\alpha'}$) by a field theory computation of tree-level exchange of supergravity massless fields between the branes. The same result can be recovered from the one-loop open string amplitudes, which have a dual interpretation in terms of tree-level exchange of closed string states between the D branes (for the cylinder) and between the D branes and O-planes (for the M\"obius strip). The setup present however some stringy features that are not captured by a pure field-theory analysis restricted to the supergravity modes. Indeed this string theory construction contains the twisted sector, and in particular the would-be scalar tachyon of \eqref{ss01}. These states do couple to branes and do mediate brane-brane interactions. Even if in the regime of interest $R \gg \sqrt{\alpha'}$, the would-be tachyonic scalar is actually very heavy and should not be kept in a low-energy effective action, its exchange is the main contribution to the brane-brane interactions at long distances that we compute below. Due to this feature, we are forced to perform the computations at the string theory level, although the results can be understood to some extent by field-theory arguments. 

The D1 branes are BPS in the superstring case with their tension equal to the R-R charge, which guarantees no interaction between them. With supersymmetry breaking turned on, branes start to interact. Indeed, consider D1 branes wrapping the Scherk-Schwarz circle (they behave like particles after compactification), at a distance $r$ in the  space coordinates. The brane-brane potential is contained in the cylinder amplitude, in a way very similar to a Casimir energy. The result of the computation is
 \be
  {\cal A}_{11}  =  \frac{1}{ \pi \sqrt{\alpha'}}   \int_0^{\infty}   \frac{d \tau_2}{\tau_2^{3/2}} e^{- \frac{ \tau_2 r^2}{4 \pi \alpha'}}  \ \left[  P_{m} 
  - P_{m+1/2}  \right]\times\frac{\theta_2^4}{2 \eta^{12}} \left(\frac{i \tau_2}{2}\right) \ .
   \label{db3}
 \ee   
 Written in the (closed string) tree-level channel, the amplitude becomes
\be
{ \tilde  {\cal A}}_{11}  =  \frac{R}{ 2 \pi \alpha'}   \int_0^{\infty}   \frac{d l}{l^4} e^{- \frac{ r^2}{2 \pi \alpha' l}}  \sum_n  \left[  1-(-1)^n  \right] \frac{\theta_4^4}{2 \eta^{12}} (i l) e^{- \pi l \frac{n^2 R^2}{2 \alpha'}} \ .       
   \label{db4}
 \ee   
Notice that only massive states contribute to the D1-D1 brane interactions (since massless states have $n=0$). In the region of interest $r, R \gg \sqrt{\alpha'}$ a standard field theory computation  does not capture the string result (\ref{db4}).  Indeed, in the region $r \gg \sqrt{\alpha'}$  the main contribution to the brane-brane interaction comes from the region $l \rightarrow \infty$ and therefore from the lightest closed string states. However, since the even winding contribution which include the supergravity states vanishes due to a cancellation between the NS-NS and the R-R sectors (this is easily seen in the winding sum in \eqref{db4}), the main contribution to the interaction comes from odd windings containing the would-be tachyon scalar in the closed string spectrum. It is illuminating to write the D1-D1 brane amplitude in a field theory format,  by recasting the potential in a way which involves an integral over the noncompact momenta of the closed strings exchanged, by using the identity
 \be
 \int_0^{\infty}   \frac{d l}{l^4} e^{- \frac{ r^2}{2 \pi \alpha' l} - \frac{ \pi l}{2}  \alpha' m_n^2} = \frac{\alpha'^3}{8 \pi} \int d^8 k \frac{e^{i {\bf k} {\bf r}}}{k^2 + m_n^2}  \ .    \label{db5}
 \ee
Then the potential is given by
\be
V_{11}  =  - \frac{  R \alpha'^2}{\pi^2}\int d^8 k  \ e^{i {\bf k} {\bf r}}  \  
 \lb \frac{1-1}{k^2} + \frac{1}{8} \ \frac{1}{k^2 + \frac{R^2}{\alpha'^2} - \frac{2}{\alpha'}}  +\dots \rb\ .   \label{db6}
\ee
 The contribution of the zero modes vanishes at one-loop, according to our computation, which is the statement that at this order the interaction of D1 branes is still governed by the BPS relation between the tree-level tension and charge $T_1=Q_1$. Indeed, since the one-loop contribution is exclusively mediated by massive states, it is short-ranged and therefore cannot be interpreted as coming from an imbalance between the tension and charge of the branes. In particular, although those short-range forces are attractive, we do not consider that as a violation of the WGC. In short, at first order those D1 branes marginally verify the WGC (understood as a constraint on the couplings to massless supergravity modes), and one needs to compute at higher loops to see the effect of SUSY breaking  on the WGC.

In a realistic compactification only four spacetime dimensions are noncompact. In this case, the brane-brane potential for $r \gg \sqrt{\alpha'}$ becomes 
 \be
  V_{11} =  - \frac{  R \alpha'^2}{8\pi^2 V_5} \sum_{\bf p} \int d^3 k  \ e^{i {\bf k} {\bf r}}  \  \frac{1 }{k^2 + m_{\bf p}^2+ \frac{R^2}{\alpha'^2}- \frac{2}{\alpha'}} \ ,  
  \label{db7}
 \ee 
 where $ \sum_{\bf p} $ is the sum over all Kaluza-Klein masses in the five additional internal dimensions. The result is particularly simple if the five additional dimensions are very small, i.e.~$R_I \ll R,r$, in which case one can neglect the corresponding massive modes contributions. In this limit (and using $R \gg \sqrt{\alpha'}$), the total potential energy is well approximated at large distances $r \gg \sqrt{\alpha'}$ by
 \be
 V_{11} \sim   \ -  \   \frac{  R \alpha'^2}{4 V_5} \frac{ e^{-r\sqrt{\frac{R^2}{\alpha'^2}-\frac{2}{\alpha'}}}}{r}    \ .  
  \label{db8}
 \ee
The interactions are dominantly due to the exchange of the would-be tachyonic scalar, as in the previous 9D case, so that the conclusion regarding the WGC is identical.

\subsection{Beyond one-loop arguments and the WGC}\label{beyondoneloop}

Now, we provide arguments in favour of the fact that our one-loop computations enable us to conclude that the WGC is verified (in a non-trivial way) in the present setup. This would naively demand higher loop computations, but actually the relevant piece of information, the quantum correction to the brane tension, is already determined by the previous computations. 

The D1 brane self-energy is obtained from \eqref{db4} by considering a single D1 brane and setting the spacetime distance ${\bf r}=0$. One gets the approximate result
 \be
{ \tilde  {\cal A}}_{11}  =  \frac{16}{ \pi^2 R}   \sum_n   \frac{1} {(2n+1)^2 } \ . 
     \label{db9} 
 \ee  
From this result, one obtains the one-loop corrected tension of the D1 brane wrapping the circle\footnote{For each D1 brane, there is also an interaction with the background D9 branes/O9-planes. It is unclear to us that this potential energy, localized on the D1 brane but Wilson line/position dependent, should be interpreted as an additional correction to the D1 brane tension. In any case, since it is of the same sign and magnitude as the self energy of the D1 brane, including it or not would not modify the qualitative features of what we discuss next.}, which can be written either as a corrected D1 brane tension or as the mass $M_0$ of the wrapped brane on the circle 
\be
T_{1, \rm eff} = T_1 -  \frac{2}{ \pi^3 R^2}  \sum_n   \frac{1} {(2n+1)^2}  \ = \ T_1 -  \frac{1} {2 \pi R^2}  \quad, \quad 
M_0 = 2 \pi R T_{1, \rm eff} 
\ ,  \label{db10}
\ee
where $T_1 = \frac{\sqrt{\pi}}{\sqrt{2}\kappa_{10}} (4 \pi^2 \alpha')$ is the standard type I D1 brane tension. Notice that this one-loop corrected tension is {\it lower} than the tree-level one, due to supersymmetry breaking (remember that $T_1\sim\mathcal{O}(g_s^{-1})$, so the correction is indeed of order $\mathcal{O}(g_s)$ with respect to the original value). 

  Let us rewrite the D1-D1 brane interactions \eqref{db6} in a slightly more general way as a contribution from the zero modes  $V_{11} ^{(0)}$  and contributions from massive states $V_{11} ^{(n)}$. The contribution of the zero-mode $V_{11} ^{(0)}$  vanishes at one-loop, according to our computation in section \ref{d1oneloop}. However, since the one-loop contribution comes exclusively from massive states, it is short-ranged and therefore any higher-order/loop correction leading to a zero-mode exchange changes dramatically the interaction at large distances. We consequently parametrize the zero-mode higher-loop contributions by introducing three parameters:  $T_{1,\rm eff}$ and $Q_{1,\rm eff}$ are the quantum corrected  brane tension and charge, whereas $m_0$ denotes the mass of the dilaton generated by quantum corrections. With these changes in mind, at large distances $r \gg \sqrt{\alpha'}$ where the main contribution comes from the lightest closed string states exchanged between the branes, we arrive at the following expression for the D1-D1 brane interaction
\be
\ba
V_{11} =  V_{11} ^{(0)}  +  V_{11} ^{(n)} \,, \quad  {\rm where } \quad  & V_{11} ^{(0)} =   \frac{ R \alpha'^2}{2\pi^2} \int d^8 k  \ e^{i {\bf k} {\bf r}} \  \left[ \frac{Q^2_{1,\rm eff}/Q_1^2}{k^2}  - \frac{T^2_{1,\rm eff}/T_1^2}{4}  \left(  \frac{1}{k^2 + m_0^2}+   \frac{3 }{k^2} \right) \  \right]  \ , \\
&  V_{11} ^{(n)} =  - \frac{  R \alpha'^2}{8\pi^2} \int d^8 k  \ e^{i {\bf k} {\bf r}}  \  \frac{1}{k^2 + \frac{ R^2}{\alpha'^2}- \frac{2}{\alpha'}} \ .   \label{wg1}
 \ea
\ee
 The zero-mode contribution can also be written in terms of the supergravity 10D Planck mass $\kappa_{10}$,
 \be
 V_{11} ^{(0)} =  16 \kappa_{10}^2 \pi R \int \frac{d^8 k}{(2\pi)^8}  \ e^{i {\bf k} {\bf r}} \  \left[ \frac{Q^2_{1,\rm eff}}{k^2}  - \frac{T^2_{1,\rm eff}}{4}  \left(  \frac{1}{k^2 + m_0^2}+   \frac{3 }{k^2} \right) \  \right]  \ .
 \ee
  In (\ref{wg1}),  the corrected tension of the wrapped D1 brane $T_{1,\rm eff}$ is defined  in (\ref{db10}) and the relative factor of 1/4 (3/4) denotes the contribution of the dilaton (graviton).  The one-loop corrected charge  $Q_{1,\rm eff}$ will be discussed below. The massive contributions $V_{11} ^{(n)}$ contain the one-loop computation performed in section \ref{d1oneloop}. 
    
Again, in a realistic compactification only four spacetime dimensions are noncompact. In this case, the brane-brane potential becomes 
 \bea
 && V_{11} ^{(0)} =\sum_{\bf p} \frac{16 \kappa_{10}^2 \pi R}{(2\pi)^8V_5}  \int d^3 k  \ e^{i {\bf k} {\bf r}} \  \Biggl[ \frac{Q^2_{1,\rm eff}}{k^2+m_{\bf p}^2}  - \frac{T^2_{1,\rm eff}}{4} \left(  \frac{1}{k^2 + m_{\bf p}^2 + m_0^2}+  
  \frac{3 }{k^2 + m_{\bf p}^2 } \right)  \Biggr]  \ , \nonumber \\ 
 && V_{11} ^{(n)} =  - \frac{  R \alpha'^2}{8\pi^2 V_5} \sum_{\bf p} \int d^3 k  \ e^{i {\bf k} {\bf r}}  \  \frac{1 }{k^2 + m_{\bf p}^2+ \frac{R^2}{\alpha'^2}- \frac{2}{\alpha'}} \ ,  
  \label{wg2}
 \eea           
 where $ \sum_{\bf p} $ is the sum over all Kaluza-Klein masses in the five additional internal dimensions. The result is particularly simple if the five additional dimensions are much smaller than $R$ and $r$, in which case one can neglect the contributions from the corresponding massive modes. In this limit, it is more transparent to express the  total potential energy in terms of the four-dimensional Planck mass $M_P$, for which the graviton exchange provides the Newton potential in terms of the mass $M_0 = 2 \pi R T_{1,\rm eff}$ and the charge $Q_0 = 2 \pi R Q_{1,\rm eff}$ of the wrapped D1 brane. In this way, one gets the approximate potential
 \be
 V_{11} \sim  \  \frac{1}{M_P^2}\left[   \ \frac{ \frac{4}{3} Q_0^2  - M_0^2  -  \frac{1}{3} M_0^2  e^{- m_0 r} }{r}  \ -  \ 
\frac{Q_0^2}{3}   \   \frac{ e^{-r\sqrt{\frac{R^2}{\alpha'^2}-\frac{2}{\alpha'}}}}{r}  \right]  \ .  
  \label{wg3}
 \ee     
 This expression is valid for distances $r \gg \sqrt{\alpha'}$, whereas for shorter distances one expects the one-loop potential to be a good approximation, which has a constant limit when $r \to 0$. 
 
 As discussed above, the correction to the D1 brane tension is generated by the massive contributions $V_{11} ^{(n)}$ between the same brane (${\bf r} = 0$), and it is negative. The correction to the charge would, on the other hand, come from a genus $3/2$ computation, which was not yet performed to our knowledge. However, a quantum correction to the R-R charge of the brane would be of the form $\int C_2 e^{\phi}$, where $\phi$ is the dilaton. Such a coupling would violate the gauge symmetry of the R-R gauge field $C_2$, which seems implausible in perturbation theory. Corrections to the R-R field kinetic terms are possible though, and this would generate a renormalization of the R-R charge. A similar correction to the dilaton kinetic term should also contribute to the renormalization of the tension. However, such corrections would arise from one loop calculations and would be associated to ${\cal O}(g_s^2)$ corrections. We thus do not expect them to dominate the one-loop contribution to the tension, which is $\mathcal{O}(g_s)$, and therefore
 \be
 T^2_{1,\rm eff} <  Q^2_{1,\rm eff}   \ \Longleftrightarrow 
 \ M_0^2 < Q_0^2 \ .    \label{wg4Bis} 
  \ee
As a consequence, at short distances the potential is attractive whereas it is repulsive at large distances.
If on the contrary the bound \eqref{wg4Bis} was violated, the potential would remain attractive also at large distances\footnote{When the dilaton is massive, $m_0 > 0$, the condition is actually $M_0^2 > \frac43 Q_0^2$.}. Our perturbative arguments dismiss such a possibility and we conclude that the weak gravity conjecture holds in our setup, and the massless modes exchange which it constrains determines the brane-brane dynamics at large distances.

\subsection{Bound states of branes}\label{boundstates}

\begin{figure}[htb]
\centering
\includegraphics[scale=0.45]{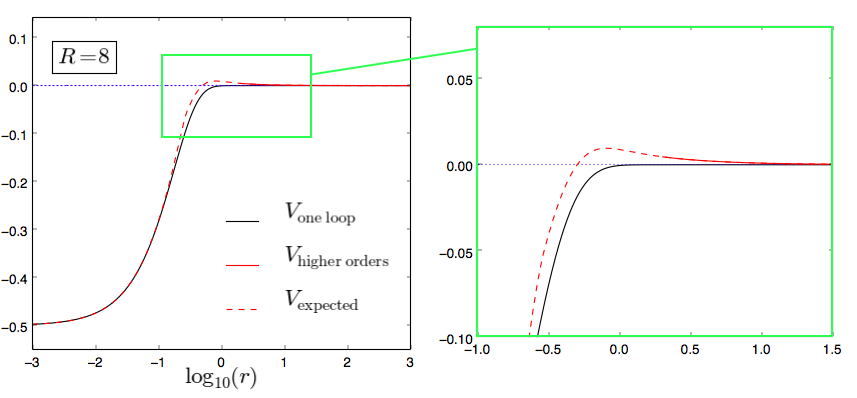}
\caption{The D1-D1 potential as a function of the distance in the transverse space (the potentials and distances are expressed in units of $\alpha'$, we fixed $R=8$, $g_s=0.2$, $V_5\sim1.5^5$)}
\label{fig:potential}
\end{figure}

An important outcome of the previous discussion is that the negative self-energy of D1 branes and the decrease in the effective brane tension also implies that it is energetically favorable to form bound states of D1 branes. 
This eventually can lead to the formation of black holes. Consequently, black holes stability arguments, which are sometimes used in discussions about the WGC, are different in the small and large distance regions. To address this question, one needs to study the regime interpolating between large distances, where higher-order effects dominate and verify the WGC as argued above, and small distances where the one-loop potential induces an attraction. Knowing the $r=0$ value of the potential given in (\ref{db9}) and its asymptotic behaviour (\ref{wg3}), we understand that it reaches a maximal value and has the shape depicted in figure \ref{fig:potential}.

To estimate the location $r_0$ of the maximum, we can use (\ref{wg3}) if $r_0$ is in its validity regime. When $m_0=0$, we obtain 
\be
r_0=-\frac{1}{\sqrt{\frac{R^2}{\alpha'^2}-\frac{2}{\alpha'}}}\left[1+W\left(8\frac{T^2_{1,\rm eff}-T_1^2}{eT_1^2}\right)\right] \approx \frac{\alpha'}{R}\log\left(\frac{R^2}{g_s\alpha'}\right) \ ,
\label{r0}
\ee
where $W$ is the Lambert $W$ function. 
This expression, obtained from (\ref{wg3}), can be trusted if $r_0\gg \sqrt{\alpha'}$, which can be rewritten as a constraint on the string coupling
\be
g_s\ll \frac{R^3}{\alpha'^{3/2}}e^{-\frac{R}{\sqrt{\alpha'}}} \ .
\label{gSBound}
\ee
In this case, black holes of size smaller than $r_0$ would be stable remnants. 
However, we expect from black hole constructions in string theory that there should only be a finite number of such remnants: 
one can guess that if the number of D1 constituents is large and the bound state size becomes or order $r_0$ or larger, repulsive forces will prevent more D1 branes to bind and therefore larger charge/mass remnants to form. Calculating this finite number of bound states is beyond the scope of this paper, but we could try to estimate it by comparing $r_0$ with the scale at which we expect the D1 branes solutions of supergravity to break down,\footnote{This scale is the one for which the harmonic function $h(r)=1+\frac{R_S}{r}$, which defines the D1 brane solution, starts to deviate significantly from one.} $r_S\sim\frac{N_1g_s\alpha'^3}{V_5}$, where $N_1$ is the number of stacked D1 branes. Using \eqref{r0}, we can derive the following estimate,
\begin{equation}\label{RSr0}
N_{\mathrm{crit}} \equiv N_1 \frac{r_0}{r_S} \approx \frac{1}{g_s} \frac{V_5}{\alpha'^{5/2}} \frac{\alpha'^{1/2}}{R} \log\left(\frac{R^2}{g_s\alpha'}\right) \,,
\end{equation}
where all D1 branes configurations with $N_1<N_{\mathrm{crit}}$ correspond to situations where the attractive force is felt even in the regime where supergravity applies.

This number becomes small in the decompactification limit $R \gg \sqrt{\alpha'}$: when the supersymmetry breaking (Scherk-Schwarz) radius goes to infinity, the one-loop potential vanishes since supersymmetry is recovered, no attraction nor repulsion remains, and the WGC, as well as the stability of black holes, is marginally retrieved. Furthermore, (\ref{RSr0}) also shows that the smaller $g_s$, the more stable bound states can exist. If $m_0\neq 0$, $r_0$ becomes smaller than (\ref{r0}) and the appearance of such states is slightly suppressed in the limit $g_s \rightarrow 0$, but the behaviour remains qualitatively the same. Such a scaling of $N_\text{crit}$ with $g_s$ is reminiscent of the swampland distance conjecture. However, the regime of validity of our model does not allow to fully test this idea. Indeed, to avoid that the state discussed in \eqref{ss01} be tachyonic, we need to ensure that $R\gtrsim\sqrt{\alpha'}$. On the other hand, given the 4D Planck mass
\begin{equation}
M^2_P \sim \frac{V_5}{\alpha'^{5/2}} \frac{R}{\alpha'^{1/2}} \frac{1}{\alpha' g_s^2} \ ,
\end{equation}
\eqref{RSr0} can be rewritten as follows,
\be
N_{\mathrm{crit}}\sim \alpha'M_P^2\underbrace{\frac{g_s\alpha'}{R^2}\log\left(\frac{g_s\alpha'}{R^2}\right)}_{\leq 1} \lesssim \frac{M_P^2}{m_\text{KK}m_\text{windings}}
\ee
(where we imposed that the string theory is weakly-coupled), so that we cannot get a large number of those bound states without an other tower of states (namely the KK or the winding modes) becoming light in Planck units. This limits our ability to identify the bound states as the states predicted by the swampland distance conjecture.

On the other hand, when the string coupling increases, $r_0$ decreases and it will eventually not be consistent to use (\ref{wg3}) and (\ref{r0}). 

\subsection{Intermediate summary}\label{summary1}

We studied a type I string setup with broken SUSY \`a la Scherk-Schwarz, and our main result is that the weak gravity conjecture (for D1 branes) holds there. This is due to a decrease of the D1 brane tension (and no comparable decrease of their gauge R-R charge), which is seen after taking the one-loop corrections from massive stringy modes into account. We reached this conclusion by defining the WGC in terms of couplings to massless fields (in the gravitational and gauge sector), disregarding the short-range attractive effects due to massive fields as a violation of the WGC. Nonetheless, those effects were crucial for the decrease of the tension, leading to long range repulsive forces. In the lower dimensional effective theory these D1 branes, wrapped around the Scherk-Schwarz circle, behave as particles charged under a $U(1)$-gauge symmetry with $Q_\mathrm{eff} > M_\mathrm{eff}$. Overall, this leads to a picture in which the weak gravity conjecture seems to be respected.

More generally, we saw as advertised that, even in the simple case we discussed, there exist already several features which are of interest for a swampland approach. First, the radius of the compact direction hosting the Scherk-Schwarz mechanism has a potential and a non-trivial dynamics, so that the model allows to study the realization of some swampland conjectures in an inherently dynamical setup. The Scherk-Schwarz radius being tied to SUSY breaking, it implies a variation of SUSY breaking effects when it evolves, which is obvious from our WGC discussion. Secondly, branes which are non-interacting in superstring setups interact when SUSY breaking is turned on, so that they verify the WGC in a non-trivial manner: quantum effects arise from the brane coupling to genuinely stringy states, and source a decrease of the brane tension consistent with the WGC. Finally, identical branes have interaction potentials whose qualitative nature evolves with the distance: they are such that the WGC repulsion dominates at large distance, whereas the contribution of massive modes kick in at smaller ones and allow for the formation of bound states. In a limit of very weak string coupling, those bound states come in large numbers and provide an unusual example of a tower of states, although we could not consistently match their behaviour with the one predicted by the distance conjecture.

\section{Magnetic fields on branes}\label{magneticSection}

We now turn to our second example, where this time SUSY breaking and branes interactions follow from turning on magnetic fields in compactified dimensions \cite{Bachas:1995ik,Blumenhagen:2000wh,Angelantonj:2000hi} (see also the section 5.11 of \cite{Angelantonj:2002ct} and references therein). Magnetic compactifications have been extensively used in particle physics models over the years (see e.g. \cite{Buchmuller:2016gib,Buchmuller:2018eog,Buchmuller:2019zhz} for recent examples), and their impact on brane dynamics can be understood from the Wess-Zumino coupling of R-R forms and gauge fields on a Dp brane \cite{Polchinski:1996na}:
\be
S_\text{WZ}\sim \int_{Dp} C \wedge e^F =\int_{Dp} \lp C_{p+1} +C_{p-1} \wedge F+\frac{1}{2}C_{p-3} \wedge F  \wedge F+...\rp \ ,
\ee
where the $C_q$ are $q$-forms from the R-R sector and $F$ is the field strength of a brane gauge field (which we took in an abelian subgroup for simplicity, since we will focus on this case later on). This expression shows that a non-trivial magnetic field along the worldvolume of a brane sources couplings to lower-than-usual degree forms\footnote{This is interpreted in terms of bound states of branes of different dimensions \cite{Witten:1995im,Douglas:1995bn}.}, and induces new dynamics for the branes. In what follows, we study this dynamics in toroidal compactifications of type II string theory by computing again (open string) one-loop partition functions, and we focus on the status of the weak gravity conjecture\footnote{The results of this section were presented by E.D. at the String Phenomenology Conference 2019, CERN, Geneva.}.

\subsection{Magnetic fields and SUSY breaking}\label{magneticSUSYbreaking}

Let us start by reviewing why magnetic fields in compact dimensions break supersymmetry from the point of view of the lower dimensional theory. In this occasion, we will encounter features relevant for later discussions.

Let us start at the level of field theory. As is well known, the momentum of charged particles encodes the minimal coupling to the gauge field, $p_\mu=\partial_\mu-iqA_\mu$. For a constant magnetic field $F=dA=$const., one can choose a gauge (called symmetric) such that $A_\mu=-\frac{1}{2}F_{\mu\nu}x^\nu$, with $x$ the spacetime coordinates. Thus, one sees that the commutation relation between momenta is modified:
\be
[p_\mu,p_\nu]=-iqF_{\mu\nu} \ ,
\ee
and the spectrum is affected too: quantizing the system leads to the appearance of the famous Landau levels \cite{LANDAU1977298}. For a six-dimensional theory (itself maybe a compactified version of a higher-dimensional theory) compactified on a torus to a four-dimensional theory with a constant magnetic field $F_{45}=F$ in the two compactified directions, the general formula for the shift of the 4D mass of the Landau levels is \cite{Bachas:1995ik}
\be
\delta M^2=p_4^2+p_5^2=(2n+1)\abs{qF}+2qF\Sigma \ ,
\ee
where $\Sigma$ is the spin operator on the torus. For instance, for a Weyl fermion in 6D, one obtains after compactifying two 4D Weyl fermions of opposite chirality for which $\Sigma=\pm\frac{1}{2}$. For a 6D gauge field, one obtains a 4D gauge field with the usual Landau level masses, i.e. $\Sigma=0$, and two real 4D scalars from the internal component with $\Sigma=\pm 1$. Thus, we see that the mass degeneracy within a 6D SUSY multiplet is broken, so SUSY must also be. This breaking is understood as being spontaneous. The multiplicity of each Landau level is $q\frac{V_{T_2}}{2\pi}F$, which is integer from the Dirac quantization condition (when the charge is also quantized).

An important feature is the presence of tachyons: indeed, the $\Sigma = -\text{sgn}(qF)$ helicity state of the internal component of a massless 6D gauge field has a negative mass $-\abs{qF}$ \cite{Nielsen:1978rm}. In string theory, such tachyonic scalars from internal components of gauge fields are generically present and this will force us to include at least two magnetic fields (this also allows to preserve part
of the original SUSY).

Since we are interested in brane dynamics in string theory, let us now turn to the open string analog of this discussion. It preserves all of the aforementioned features and goes as follows: the worldsheet theory of open strings is modified by the presence of magnetic fields, since the end of open strings couple to gauge fields, as seen from the bosonic part of the action (here in conformal gauge, see \cite{Bachas:1995ik} for the full superstring treatment):
\be
S_\text{bosonic string}=-\frac{1}{4\pi\alpha'}\int d^2\sigma (\partial X)^2-q_L\int d\tau \ A_\mu\partial_\tau X^\mu\Big|_{\sigma=0}-q_R\int d\tau \ A_\mu\partial_\tau X^\mu\Big|_{\sigma=\pi} \ .
\label{stringMagneticAction}
\ee
The equations of motion do not feel the endpoints couplings and are still describing free propagating fields, but the boundary conditions are affected, and read\footnote{Those boundary conditions have a nice interpretation in terms of rotated branes \cite{Berkooz:1996km}, which is seen by T-dualizing along one of the directions carrying the magnetic field.}
\be
\sigma=0: \ \partial_\sigma X^\mu - 2\pi\alpha'q_LF^\mu{}_\nu\partial_\tau X^\nu=0 \ , \quad \sigma=\pi: \ \partial_\sigma X^\mu + 2\pi\alpha'q_RF^\mu{}_\nu\partial_\tau X^\nu=0 \ .
\ee
Let us as before turn on a magnetic field $F_{45}=F$ for two internal dimensions. The mode expansions of the string coordinates are \cite{Abouelsaood:1986gd}:
\be
Z(\tau,\sigma)=\frac{X^4+iX^5}{\sqrt 2}=z+i\lp\sum_{n=1}^\infty a_n\psi_n(\tau,\sigma)-\sum_{n=0}^\infty b^\dagger_n\psi_{-n}(\tau,\sigma) \rp \ ,
\ee
with 
\be
\psi(\tau,\sigma)=\frac{1}{\sqrt{\abs{n-\epsilon}}}\cos((n-\epsilon)\sigma+\gamma)e^{-i(n-\epsilon)\tau} \ , \text{with } \bigg\{\begin{matrix}\tan\gamma=2\pi\alpha' q_L F{\color{white}+12000\arctan(2\pi\alpha' q_RF)} \\ \pi\epsilon=\arctan(2\pi\alpha' q_LF)+\arctan(2\pi\alpha' q_RF)\end{matrix} \ ,
\label{defModeMagnetic}
\ee
(we always choose $0\leq\epsilon\leq1$ in what follows). The spectrum of the string follows, and it is given by
\be
\delta M^2=\frac{1}{2\alpha'}\Big[(2n+1)\abs{\epsilon}+2\epsilon\Sigma\Big] \ ,
\ee
(with a degeneracy $(q_L+q_R)\frac{V_{T_2}}{2\pi}F$) on top of the usual contributions of the other coordinates and of the string excitations. It is very analogous to the field theory expression, in particular there are tachyons (for instance from internal gauge fields).

Tachyons can be removed by introducing two magnetic fields, for instance $F_{45}=F_1,F_{67}=F_2$ (along the same generator of the gauge group) in four internal dimensions. 
In this case, the Landau level spectrum becomes
\be
\delta M^2=\frac{1}{2\alpha'}\bigg[\sum_{i=1,2}(2n_i+1)\abs{\epsilon_i}+2\sum_{i=1,2}\epsilon_i\Sigma_i\bigg] \ .
\ee
This expression leaves the possibility to remove the tachyons from the spectrum and maintain (half of) SUSY. This is achieved when $F=\pm \star F$, depending on the sign of $F_1,F_2$. However, we see that the two options are either the presence of tachyons with SUSY breaking, or a SUSY preserving tachyon free spectrum. We need one more step to get SUSY breaking without tachyons, which is achieved by introducing a third magnetic field, as we discuss now.

\subsection{Tachyons and the WGC}\label{tachyonsWGC}

We now study brane dynamics in a specific model, namely type IIB string theory with a compactified space containing magnetized D9 branes wrapping three 2D tori, equivalent after
T-dualities to IIA string theory with rotated D6 branes. We consider therefore D6 branes wrapped around the internal space (thus particles with respect to 4D spacetime),
\be
\begin{array}{c|cccc|cccccc}
\text{Coord. axis}&0&1&2&3&4&5&6&7&8&9\\
\hline
\text{ Rotation/Magnetic field }&&&&&\multicolumn{2}{c}{F_1}&\multicolumn{2}{c}{F_2}&\multicolumn{2}{c}{F_3}\\
\text{D6}&\times&&&&\times&\times&\times&\times&\times&\times
\end{array}
\ee
and we compute the interaction potential between two such D6 branes separated by a distance $r$ in 4D spacetime. At one-loop and using open-closed duality, this amounts to computing an annulus vacuum diagram (remember $\cal V = -A$), which reads (up to an irrelevant factor involving $\pi$s and $\alpha'$ - we keep carefully the signs though since they are the key point here)
\be
{\cal A}_{D6-D6}=q^3k_1k_2k_3\int\frac{d\tau_2}{\tau_2^{3/2}}\frac{\sum_{\alpha,\beta}\eta_{\alpha\beta}\theta[\begin{smallmatrix}\alpha\\\beta\end{smallmatrix}](0|\tau)\prod_{i=1}^3 \theta[\begin{smallmatrix}\alpha\\\beta\end{smallmatrix}](\epsilon_i\tau|\tau)}{2\eta^6(\tau)}\frac{(2i\eta)^3(\tau)}{\prod_{i=1}^3 \theta_1(\epsilon_i\tau|\tau)}\bigg|_{\tau=\frac{i\tau_2}{2}}\times e^{-\frac{\tau_2}{4\pi\alpha'}r^2}
\label{annulusLoopChannel}
\ee
where $\eta_{\alpha\beta}= (1,-1,-1,-1)$ encode the GSO projection phases in the NS and R sectors, $q=q_L+q_R$ is the overall charge of the corresponding open string sector under the gauge group generator 
hosting the magnetic field (see \eqref{stringMagneticAction}), the $\epsilon_i$s are given in \eqref{defModeMagnetic} and the $k_i$s are the integer numbers $\frac{V_{T_{2,i}}}{2\pi}F_i$, so that the factor in front of the integral accounts for the degeneracy of the Landau levels. Performing a modular S-transformation and using the Jacobi identity, we can write this amplitude as follows
\be
{\cal A}_{D6-D6}=-q^3k_1k_2k_3\int\frac{dl}{l^{3/2}}\frac{\theta_1(\frac{\epsilon_1+\epsilon_2+\epsilon_3}{2}|il)\theta_1(\frac{-\epsilon_1+\epsilon_2+\epsilon_3}{2}|il)\theta_1(\frac{\epsilon_1-\epsilon_2+\epsilon_3}{2}|il)\theta_1(\frac{\epsilon_1+\epsilon_2-\epsilon_3}{2}|il)}{\eta^3(il)\prod_i\theta_1(\epsilon_i|il)}e^{-\frac{r^2}{2\pi\alpha'l}} \ .
\label{annulusTreeChannel}
\ee
From this expression, it is possible to see that if one of the magnetic field is zero, let's say the first one for definiteness so that $\epsilon_1=0$, the brane potential is attractive if $\epsilon_2\neq \pm\epsilon_3$, in which case there are also tachyons in the spectrum\footnote{The spectrum can be read off from the amplitude \eqref{annulusLoopChannel} at $r=0$ by performing a $q=e^{2i\pi\tau}$ expansion. The masses of the physical particles in the spectrum appear as follows:
\be
{\cal A}=\int\frac{d\tau_2}{\tau_2^{3/2}}\sum_{\text{masses } M}\sum_\text{spins}\text{degeneracy}\times\text{spin d.o.f}\times e^{-\pi\alpha'\tau_2M^2} \ ,
\ee
up to an overall coefficient.}. If $\epsilon_2= \pm\epsilon_3$, the tachyons vanish and so does the potential. This is nothing but the correlation between SUSY and the presence of tachyons with two magnetic fields that we announced before. Things get interesting when keeping all three magnetic fields. This allows for broken SUSY and tachyon-free spectra. Indeed, in such a case, the would-be tachyons from the internal gauge field components have masses
\be
\alpha' M^2_1=1-\frac{\epsilon_1+\epsilon_2+\epsilon_3}{2} \ , \quad \alpha' M^2_2=\frac{-\epsilon_1+\epsilon_2+\epsilon_3}{2} \ , \quad \alpha' M^2_3=\frac{\epsilon_1-\epsilon_2+\epsilon_3}{2} \ , 
\quad \alpha' M^2_4=\frac{\epsilon_1+\epsilon_2-\epsilon_3}{2} \ .
\ee
Tachyons are avoided if "triangle inequalities" are respected \cite{Aldazabal:2000dg,Ibanez:2001nd}:
\be
\epsilon_1+\epsilon_2\geq\epsilon_3 \ , \quad \epsilon_1+\epsilon_3\geq\epsilon_2 \ , \quad \epsilon_2+\epsilon_3\geq\epsilon_1 \ , \quad \epsilon_1+\epsilon_2+\epsilon_3\leq 2 \ , \quad 
\ee
which can be verified even when SUSY is broken. Nevertheless, these conditions tell us a lot about the brane dynamics, since they explicitly appear in the large $r$ approximation of  \eqref{annulusTreeChannel}\footnote{This large $r$ approximation expression (\ref{mag1}) is valid for $\epsilon_1 + \epsilon_2+\epsilon_3 \leq 2$.},
\be
\ba
{\cal A}_{D6-D6}=\bigg|_{r\gg\sqrt{\alpha'}}&-q^3k_1k_2k_3\times\\
&\int\frac{dl}{l^{3/2}}\frac{\sin(\frac{\pi(\epsilon_1+\epsilon_2+\epsilon_3)}{2})\sin(\frac{\pi(-\epsilon_1+\epsilon_2+\epsilon_3)}{2})\sin(\frac{\pi(\epsilon_1-\epsilon_2+\epsilon_3)}{2})\sin(\frac{\pi(\epsilon_1+\epsilon_2-\epsilon_3)}{2})}{\prod_i\sin(\pi\epsilon_i)}e^{-\frac{r^2}{2\pi\alpha'l}} \ . \label{mag1}
\ea
\ee
From this expression one can conclude that {\it branes repel whenever there are no tachyons in the spectrum} (and they attract when there are: indeed, one can show that $M_i^2+M_j^2\geq 0$ $\forall i\neq j$ so that there is at most one tachyon - recall that we chose $0\leq\epsilon_i\leq1$). We find again that SUSY breaking generates repulsive long-range brane interactions, as predicted by the weak gravity conjecture. This time, the repulsion is a genuine one-loop effect, and one can identify the contributions of NS-NS and R-R modes to clearly see the mismatch between charge and tension. Indeed, one can rewrite the integrand as follows
\be
\ba
q^3k_1k_2k_3\frac{\sin(\frac{\pi(\epsilon_1+\epsilon_2+\epsilon_3)}{2})\sin(\frac{\pi(-\epsilon_1+\epsilon_2+\epsilon_3)}{2})\sin(\frac{\pi(\epsilon_1-\epsilon_2+\epsilon_3)}{2})\sin(\frac{\pi(\epsilon_1+\epsilon_2-\epsilon_3)}{2})}{\prod_i\sin(\pi\epsilon_i)}\qquad\qquad&\\
\sim\frac{q^3k_1k_2k_3}{\prod_i\sin(\pi\epsilon_i)}\bigg(\underbrace{1+\sum_i\cos(2\pi\epsilon_i)}_\text{NS-NS}-\underbrace{4\sum_i\cos(\pi\epsilon_i)}_\text{R-R}\bigg)& \ .
\ea
\ee

\begin{figure}[htb]
\centering
\includegraphics[scale=0.45]{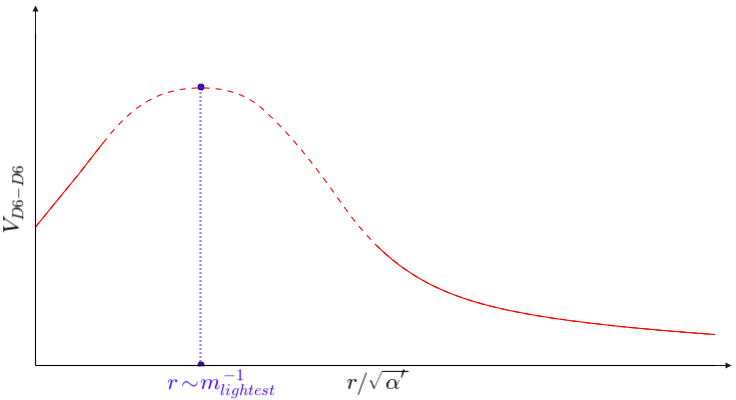}\\
\includegraphics[scale=0.45]{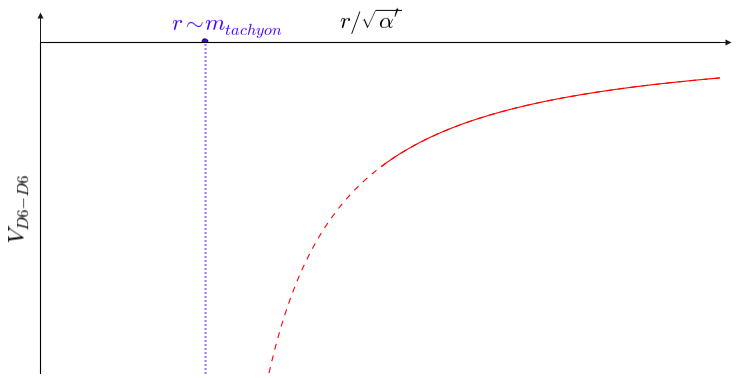}
\caption{Schematic shape of the D6-D6 potential as a function of the distance in 3D space, for a tachyon-free theory (upper panel - $m_{lightest}$ refers to the lightest massive string state contributing to the potential) or when a tachyon is present (lower panel)}
\label{fig:potentialMagnetic}
\end{figure}

Let us end this discussion by again asking the question of bound states in this theory. To answer this, one needs the full shape of the potential, similarly to figure \ref{fig:potential} (although here the potential is fully one-loop and explicitly computed). We presented the large distance part of it previously, and one can also get a reasonably simple expression for small spacetime separations between the branes ($r\ll\sqrt{\alpha'}$):
\be
\ba
{\cal A}_{D6-D6}=&\bigg|_{r\ll\sqrt{\alpha'}}-q^3k_1k_2k_3\times\\
\int\frac{d\tau_2}{\tau_2^{3/2}}&\frac{\sinh(\frac{\pi \tau_2(\epsilon_1+\epsilon_2+\epsilon_3)}{4})\sinh(\frac{\pi \tau_2(-\epsilon_1+\epsilon_2+\epsilon_3)}{4})\sinh(\frac{\pi \tau_2(\epsilon_1-\epsilon_2+\epsilon_3)}{4})\sinh(\frac{\pi \tau_2(\epsilon_1+\epsilon_2-\epsilon_3)}{4})}{\prod_i\sinh(\frac{\pi \tau_2\epsilon_i}{2})}e^{-\frac{\tau_2r^2}{4\pi\alpha'}} \ .
\ea
\ee
There are two potentially troublesome regions in the $\tau_2$ integral. There is no divergence close to $\tau_2=0$,  since the $\sinh$s scale linearly with $\tau_2$ there. However, the $\tau_2=\infty$ region is potentially problematic. When there are no tachyons, the arguments of the $\sinh$s are all positive. Thus, the integrand approximately scales as $e^{-\frac{\tau_2r^2}{4\pi\alpha'}}/\tau_2^{3/2}$ and the integral converges whatever $r$. If there is a tachyon, say $M_2^2<0$ for definiteness, the scaling becomes $e^{-\tau_2\lp\frac{r^2}{4\pi\alpha'}+\alpha'M_2^2\rp}/\tau_2^{3/2}$ and the integral diverges when $r\lesssim\abs{M_2}$ in string units. Eventually, the potential looks as in figure \ref{fig:potentialMagnetic}.
It may be interesting to know how to connect a discussion of bound states with such potentials to the existence of black holes in type IIA string theory. We leave this for future work. 

\subsection{Intermediate summary}\label{summary2}

In this second section, we discussed how supersymmetry breaking through magnetic fields along compact directions could affect the weak gravity conjecture. Specifically, we computed the long-range forces between D6 branes in type IIA string theory and showed that whenever the open string spectrum has no tachyons, the branes self-repel. This is consistent with the WGC. Furthermore, we discussed the precise shape of the brane potential (including the contribution of all the string modes).  

\section{Conclusion}\label{summary}

Supersymmetry breaking in string theory comes together with a number of ingredients relevant for studies of the swampland. In particular, those include runaway potentials for moduli fields and brane interactions. Thus, in light of the non-supersymmetric nature of the observable world, a precise assessment of swampland conjectures in string setups with broken SUSY is important.

We made a step in this direction by studying the weak gravity conjecture for D branes when SUSY is broken by compactification or by the introduction of magnetic fields in compact dimensions. We saw already at the level of very simple models that the way the WGC holds relies on non-trivial correlations (e.g. with the presence of tachyons when it is not satisfied) and on stringy effects (such as the quantum corrections to brane tensions generated by heavy string states). This makes us hope that interesting phenomena are yet to be discovered along these lines. For instance, it would be interesting to look at the WGC in models of brane supersymmetry breaking \cite{Antoniadis:1999xk,Aldazabal:1999jr}. We leave this for future work.

\section*{Acknowledgments}

Q.B. is supported by the Deutsche Forschungsgemeinschaft under Germany's Excellence Strategy  EXC 2121 ``Quantum Universe" - 390833306. This work was supported in part by  the ANR grant Black-dS-String ANR-16-CE31-0004-01. 

\bibliographystyle{JHEP}         
\bibliography{Corfu.bib}

\end{document}